\documentstyle[12pt,aps]{revtex}

\include{psfig}

\input epsf

\begin{document}
\title{
\hfill\parbox[t]{2in}{\rm\small\baselineskip 14pt
{~~~~~JLAB-THY-99-20}\vfill~}
\vskip 2cm
Critique of a Pion Exchange Model for Interquark Forces
}

\vskip 1.0cm

\author{Nathan Isgur}
\address{Jefferson Lab, 12000 Jefferson Avenue,
Newport News, Virginia  23606}
\maketitle

\vspace{6.0 cm}
 
\begin{abstract}

		I describe four serious defects of a widely discussed pion exchange
model for interquark forces:  it doesn't solve the ``spin-orbit problem" as
advertised, it fails to describe the internal structure 
of baryon resonances, it leads
to disastrous conclusions when extended to mesons, and it is not
reasonably connected to the physics of heavy-light systems.

\bigskip\bigskip\bigskip\bigskip

\end{abstract}
\pacs{}
\newpage

\section {Introduction}
\medskip

The idea that the low-energy degrees of freedom of QCD are quarks, gluons,
and Goldstone bosons is an old and interesting one.  In one form or
another, it has been used in a wide variety of models for the last two
decades \cite{GR,otherOPE}.
In this paper I offer a critique of a recent and widely discussed variant of
such models due to Glozman and Riska \cite{GR} in which it is proposed that
baryon spectroscopy be described by discarding the standard
one-gluon-exchange (OGE) forces of De R\'ujula, Georgi, and Glashow \cite{DGG}
(which were applied to baryons most extensively by Isgur and Karl \cite{IK})
in favor of the exchange between quarks of the octet of pseudoscalar mesons
(OPE).  I will avoid the distraction of criticizing either
practical details of the Glozman-Riska model \cite{GR} or its
theoretical foundations. I will instead
accept the original model at face value and describe what I see as
its four most serious defects. Some of these defects have been pointed out
less formally in the past, and partly in response the original OPE
model \cite{GR} has been elaborated \cite{update,fixcentral,addSV,addrho}.
Although these elaborations have not overcome the defects I describe here \cite{Glozmanreply},
I will comment upon them as appropriate in what follows.

\section {A Catalogue of Criticisms}

\subsection {The Spin-Orbit Problem is Not Solved}

	One of the central motivations for the Glozman-Riska model was to
solve the ``baryon spin-orbit  problem".   The Isgur-Karl model \cite {IK}
discards spin-orbit forces in view of the data which demand that such
forces be small, so that many of the successes of that model are due to the
OGE-induced hyperfine interactions (of both the spin-spin and tensor
types).  The authors of Ref. \cite{GR} note
that OPE produces hyperfine interactions {\it without} spin-orbit
interactions, and argue that this supports their hypothesis 
that OPE is the true origin of the residual
interquark forces ({\it i.e.,} of interactions beyond those which produce
confinement).

	This argument has a fundamental flaw.  The zeroth-order confining
potential, whose eigenstates are the basis for first order perturbation
theory in
both the Isgur-Karl and Glozman-Riska models, will produce very strong
spin-orbit forces through Thomas precession, a purely kinematic effect.
From the observed spectrum of states, it is impossible to escape the
conclusion that this source of spin-orbit forces alone would produce {\it
inverted}
spin-orbit multiplets with splittings of {\it hundreds of MeV}.  Thus the
true nature of the  ``spin-orbit problem" seems to have been misunderstood:  
it  is to
arrange a sufficiently precise {\it cancellation} between dynamically
generated spin-orbit forces and the inevitable Thomas-precession-induced
spin-orbit forces.

	These issues are  discussed in the original Isgur-Karl
papers \cite {IK}, but especially in view of some recent developments in
the subject, I will review the main points here.  For reasons that will
soon become apparent, it is best to start the discussion in the meson
sector.  The mesons also have a ``spin-orbit  problem" as can be seen by
examining the first band of positive parity excited mesons:  the four
P-wave mesons of every flavor are nearly degenerate.  Most of the observed
small non-degeneracies are due to hyperfine interactions, but the
spin-orbit matrix elements can be extracted.  For example,
by taking the isovector meson combination 
${5 \over 12}m_{a_2}-{1 \over 4}m_{a_1}-{1 \over 6}m_{a_0}$ 
one can isolate their spin-orbit matrix
element of $-3 \pm 20$ MeV.  As in the baryons, this matrix element
is much smaller than would be obtained from  OGE.
However, as already explained, this is not the point.  In fact, a
substantial ``normal" spin-orbit matrix element is {\it needed} to
cancel the strong ``inverted" spin-orbit matrix element from
Thomas precession in the confining potential \cite{LSgone}.  For
example, a recent fit \cite{LSInversion}
to the data on heavy-light mesons, including as a
limiting case the light-light  isovector mesons, gives an OGE spin-orbit
matrix element of +240 MeV and a Thomas precession spin-orbit matrix
element from the confining potential 
of -200 MeV:  both are very large but they are nearly perfectly
cancelling.

	The physics behind this cancellation has received support recently
from analyses of heavy quarkonia, where both analytic techniques
\cite{Brambilla} and numerical studies using lattice QCD \cite{latticeLS}
have shown that the confining forces are spin-independent {\it apart} from
the inevitable spin-orbit pseudoforce due to Thomas precession.  Moreover,
as has been known for more than ten years, the data on charmonia  require a
negative spin-orbit matrix element from Thomas precession in the confining potential
to cancel part of
the strength of the positive OGE matrix element.  If the charm quark were
sufficiently massive, its low-lying spectrum would be rigorously dominated
by one gluon exchange.
Indeed, one observes that the $\Upsilon$ system is closer to this ideal, as
expected.  Conversely, as one moves from $c \bar c$ to lighter quarks, the $\ell=1$
wave functions move farther out into the confining potential and the
relative strength of the Thomas precession term grows.  It is thus very
natural to expect a strong cancellation in light quark systems, though the
observed nearly perfect cancellation must be viewed as accidental.

    	With these points in mind, let us now turn to baryons.  As shown
in the original Isgur-Karl paper on the P-wave baryons \cite{IK}, 
a very similar cancellation can occur at the two-body level
in baryons.    However, unlike mesons,
baryons can also experience three-body spin orbit forces \cite{3bodyso} ({\it e.g.,}
potentials
proportional to $(\vec S_1-\vec S_2) \cdot (\vec r_1-\vec r_2) \times \vec
p_3$
where $\vec S_i$, $\vec r_i$, $\vec p_i$ are the spin, position, and momentum
of quark $i$).  The
matrix elements of these three body spin-orbit forces are all calculated
in Ref. \cite {IK}, but no apparent cancellation amongst them is found.  {\it I.e.,} the
spin-orbit  problem might more properly be called the ``baryon
three-body spin orbit problem".  In view of the facts that
one could understand the smallness of spin-orbit forces in mesons
and that the data clearly called for small
spin-orbit forces in baryons,
the Isgur-Karl model anticipated a solution to
the baryon three-body spin-orbit problem  and {\it
as a
first approximation} discarded all spin-orbit forces.
It was assumed that, as in mesons, a more precise and broadly applicable
description would have to treat residual spin-orbit interactions \cite{CapstickIsgur}.

	It should now be clear that replacing OGE by OPE is {\it not} a
step forward, but rather a step backward, in dealing with the observed
smallness of spin-orbit forces in baryons.  By eliminating the OGE
spin-orbit forces the Glozman-Riska model has not solved the baryon
three-body spin-orbit problem but
it has fully exposed ({\it i.e.,} left completely uncancelled) the strong
Thomas
precession forces from confinement.  Thus, this model has
escalated the baryon spin-orbit problem into a ``baryon two- and three-body
spin-orbit problem", not solved it. The recent extensions of the 
Glozman-Riska model to include
vector meson exchange \cite{update,addrho} have attempted to address
this problem. It is found that $\rho$ exchange can produce a very
strong spin-orbit interaction which might cancel with Thomas precession.
Needless to say, at the best such an arrangement offers no
improvement over the original situation \cite{SimulaLS}!
Since, for reasons to be described below, it would require 
independent solutions of the meson and baryon spin-orbit puzzles,
I consider this aspect of even the elaborated Glozman-Riska model to be a decisive
step backward.

\subsection {Baryon Internal Wave Functions are Wrong}

	In a complex system like the baryon resonances, predicting the
spectrum of states is not a very stringent test of a model.  The
prototypical example (and the first case in $N^*$ spectroscopy where this
issue arises) is the two $N^*{1\over 2}^-$ states found in the 1500-1700 MeV range.
In any reasonable valence quark model, two $N^*{1\over 2}^-$ states will be predicted
in this mass range: the excitation of a unit of orbital angular momentum
will create the negative parity and cost about 500 MeV in excitation energy
relative to the $N-\Delta$ center-of-mass position at 1100 MeV ({\it c.f.} the
$a_2-\rho$ splitting), and totally antisymmetric states
with overall angular momentum  ${1 \over 2}$ can be formed
by coupling  either quark spin ${3 \over 2}$ or quark spin ${1 \over 2}$ with $\ell=1$.  
In the
general case such a model will therefore give
\begin {eqnarray}
\vert N^*{1\over 2}^- (upper) \rangle &=& cos \theta_{{1\over 2}^-} \vert ^4P_N \rangle
+sin \theta_{{1\over 2}^-} \vert ^2P_N \rangle \\
\vert N^*{1\over 2}^- (lower) \rangle &=& cos \theta_{{1\over 2}^-} \vert ^2P_N \rangle
-sin \theta_{{1\over 2}^-} \vert ^4P_N \rangle
\label{eq:nhalfmixing}
\end {eqnarray}
in an obvious notation.  Since the masses of these resonances are only
known (and currently interpretable) to roughly 50 MeV, it is not extremely difficult
to arrange for a model to give a satisfactory description of the $N^*{1\over 2}^-$
spectrum.  However, among models which perfectly describe the spectrum
there is still a continuous infinity of predictions for the internal
composition of these two states since all values of $\theta_{{1\over 2}^-}$ from 0 to $\pi$
correspond to distinct states.

	It had been appreciated for some time \cite{HLC} that the peculiar
decay properties of the $N^*{1\over 2}^-$ states, and in particular the dominance of
the $N \eta$ decay of the lower state despite its phase space suppression
relative to $N \pi$, required that $\theta_{{1\over 2}^-} \simeq -35^{\circ}$.  
One of the early successes of the
Isgur-Karl model was that it makes the {\it parameter free} prediction 
$\theta_{{1\over 2}^-} = -arctan ({{\sqrt 5-1} \over 2}) \simeq -32^{\circ}$!
The quark model 
also predicts a pair of $N^*{3\over 2}^-$ states which have an analogous mixing
angle $\theta_{{3\over 2}^-}$.  The empirically determined value of 
that angle was $\theta_{{3\over 2}^-} \simeq +10^{\circ}$,
while the Isgur-Karl model predicts 
$\theta_{{3\over 2}^-} = arctan ({{\sqrt{10}} \over {14+\sqrt{206}}})
\simeq +6^{\circ}$.  In contrast,
though the OPE model produces a very acceptable 
negative parity $N^*$ spectrum, it predicts 
$\theta_{{1\over 2}^-} =  \pm 13^{\circ}$
and 
$\theta_{{3\over 2}^-} =  \pm 8^{\circ}$. Even though Ref. \cite{GR} only quotes the probabilities
of $\vert ^4P_N \rangle$ admixtures so that the critical signs of these mixing angles
are not available, these results are sufficient for one to see that
{\it  the internal structure of the predicted
states is wrong}. In concrete terms, such a $\theta_{{1\over 2}^-}$, even if it has the right sign,
will have almost no impact on explaining
the anomalously large $N \eta$ branching ratio of the
$N^*(1535){1\over 2}^-$ and the anomalously small
$N \eta$ branching ratio of the
$N^*(1650){1\over 2}^-$.

	More extreme cases of the importance of using the internal
structure of states and not just spectroscopy as tests of dynamics are
found in the positive parity band of excited baryons in the 1700-2000 MeV
range.  For example, the valence quark model 
predicts {\it five} $N^*{3\over 2}^+$ states in this
range, but only one is known.  Given that the masses of the $N^*$'s are rarely
known to better than 50 MeV, it would be an unlucky modeller who couldn't
identify one of their five predicted states with the observed state and
claim spectroscopic success!  What is far less trivial, as in the $N^*{1\over 2}^-$
sector, is to ask whether the one ``predicted" $N^*{3\over 2}^+$ state has production
and decay amplitudes consistent with the observed state, and, equally
important, to understand why the other four $N^*{3\over 2}^+$ states ``did not bark in
the night".  This is part of the well-known missing resonance problem and, as
shown in Ref. \cite{KoniukI}, the OGE mechanism of the Isgur-Karl model
provides a remarkably complete explanation across the
entire baryon spectrum for which states should have been
seen {\it and} where they are seen \cite{3/2photo}.  There is no 
evidence that the OPE
model has this critical property.

	Our discussion of the 
definitive role of internal structure 
would be incomplete without an example which touches back on
the issue of spin-orbit forces.  That the $\Lambda (1405){1\over 2}^-$ and $\Lambda
(1520){3\over 2}^-$ are not degenerate seems to be a sign that the approximation of
neglecting spin-orbit forces is imperfect.  Indeed, the discrepancy between
the Isgur-Karl model prediction of 1490 MeV and the observed mass of 1405
MeV for the lightest $\Lambda {1\over 2}^-$ state is one of the model's worst
spectroscopic failures.  
This has led to speculation that the $\Lambda (1405){1\over 2}^-$
is a $\bar KN$ bound state.
However, there is little doubt that, while its mass
is off by 85 MeV, the predicted state is to be identified with the 
$\Lambda (1405){1\over 2}^-$.  An analysis \cite{HLC} of the production and decay
amplitudes of the three expected $\Lambda {1\over 2}^-$ baryons gives a best fit with
\begin {eqnarray}
\vert \Lambda (1405){1\over 2}^- \rangle &=& 
+0.80 \vert ^2 \Lambda_1 \rangle
+0.60 \vert ^2 \Lambda_8 \rangle 
-0.04 \vert ^4 \Lambda_8 \rangle \\
\vert \Lambda (1670){1\over 2}^- \rangle &=& 
-0.44 \vert ^2 \Lambda_1 \rangle
+0.63 \vert ^2 \Lambda_8 \rangle 
+0.64 \vert ^4 \Lambda_8 \rangle \\
\vert \Lambda (1775){1\over 2}^- \rangle &=& 
+0.41 \vert ^2 \Lambda_1 \rangle
-0.49 \vert ^2 \Lambda_8 \rangle 
+0.77 \vert ^4 \Lambda_8 \rangle 
\label{eq:lambdamixing}
\end {eqnarray}
where $^2\Lambda_1$ is the quark spin ${1 \over 2}$ SU(3) singlet $\Lambda$  and
$^2\Lambda_8$ and $^4\Lambda_8$ are the quark spin ${1 \over 2}$ and ${3 \over 2}$ SU(3) octet
$\Lambda$'s, respectively.  The Isgur-Karl model gives (again with no
parameters)
\begin {eqnarray}
\vert \Lambda (1490){1\over 2}^- \rangle &=& 
+0.90 \vert ^2 \Lambda_1 \rangle
+0.43 \vert ^2 \Lambda_8 \rangle 
+0.06 \vert ^4 \Lambda_8 \rangle \\
\vert \Lambda (1650){1\over 2}^- \rangle &=& 
-0.39 \vert ^2 \Lambda_1 \rangle
+0.75 \vert ^2 \Lambda_8 \rangle 
+0.53 \vert ^4 \Lambda_8 \rangle \\
\vert \Lambda (1800){1\over 2}^- \rangle &=& 
+0.18 \vert ^2 \Lambda_1 \rangle
-0.50 \vert ^2 \Lambda_8 \rangle 
+0.85 \vert ^4 \Lambda_8 \rangle 
\label{eq:lambdamixing}
\end {eqnarray}
which is imperfect, but quite acceptable given the uncertainties in the data
and in its interpretation.  

    Similarly, the decay analyses also indicate that the $\Lambda (1520){3\over 2}^-$ is
to be identified with the lightest $\Lambda {3\over 2}^-$ of
the Isgur-Karl model. Experiment therefore tells us that,
despite the spectroscopic discrepancies, the
$\Lambda (1405){1\over 2}^-$
and $\Lambda (1520){3\over 2}^-$
are indeed spin-orbit partners which will evolve (as $m_s$ increases to $m_c$ and then to
the heavy quark limit $m_Q=\infty$) into the degenerate partners of a heavy quark symmetry spin
multiplet \cite{IW}. This fact will play an important role in Section D below.

    It has been suggested \cite{addSV} that the extension of the Glozman-Riska model 
to include other meson exchanges will correct the failure
of the OPE model to describe the internal structure of the baryon
resonances. This may be, but it remains to be demonstrated.
It has also been claimed that recent phenomenological
analyses of the  negative parity baryons \cite{Lebed,Collins}
are incompatible with the Isgur-Karl model, but support the Glozman-Riska model.
Since the Isgur-Karl model fits the data reasonably well, this can hardly
be true! In fact, for the reasons described in this Section,
it is clear that the phenomenological matrix
elements deduced by these analyses must be inconsistent
with OPE. What both analyses do show is that a {\it generic}
flavor-exchange interaction can produce a better fit 
to the data (with its {\it experimental} errors)
than a generic flavor-independent model. Given the physics that is
currently being ignored in such 
models (i.e., the unknown {\it theoretical} errors), the significance of this
observation is unclear.

\subsection {Mesons Are a Disaster}

     There are two ways in which the OPE model is a disaster for mesons:  it doesn't
produce spin-dependent interactions where they are needed and so requires that
we invoke {\it independent} mechanisms for creating splittings in mesons
and baryons, and it predicts the existence of effects in mesons which are
ruled out experimentally.

	We know from quenched lattice QCD that at least the bulk of both
meson and baryon hyperfine  interactions occur in the quenched
approximation  ({\it i.e.,} in the absence of closed $q \bar q$ loops) \cite{LatticeHFI}.  
Figures 1 show a Z-graph-induced meson exchange between quarks that
arises in the quenched approximation and could therefore in principle be 
the origin of the OPE-induced hyperfine interactions 
posited in the Glozman-Riska model.
The first problem I wish to highlight is that this mechanism can only
operate between two quarks and  {\it not} between a quark and an antiquark,
so if baryon spin-dependent interactions are dominated by OPE, meson and baryon
spin-dependent interactions must have totally different physical origins.  This
is not only unaesthetic:  as we shall see, it is also very difficult
to arrange.

%
%
\begin{center}
~
\epsfxsize=2.0in  \epsfbox{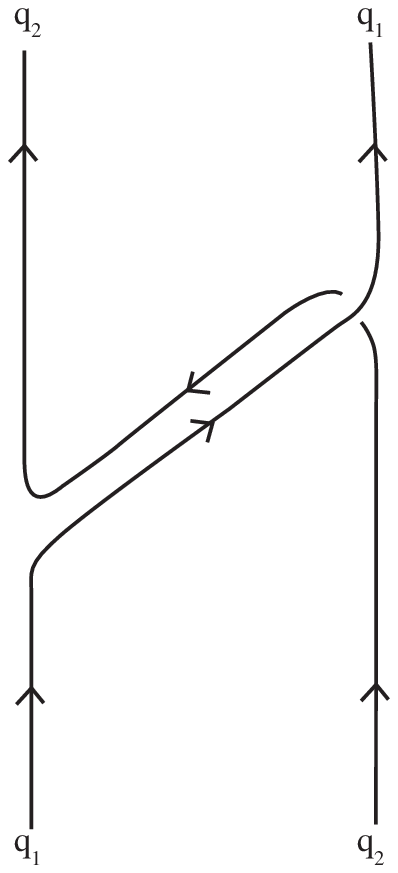}
\vspace*{0.1in}
~
\end{center}

\centerline{Fig. 1(a):  Z-graph-induced meson exchange between two quarks.}

\bigskip

%
%
\begin{center}
~
\epsfxsize=6.5in  \epsfbox{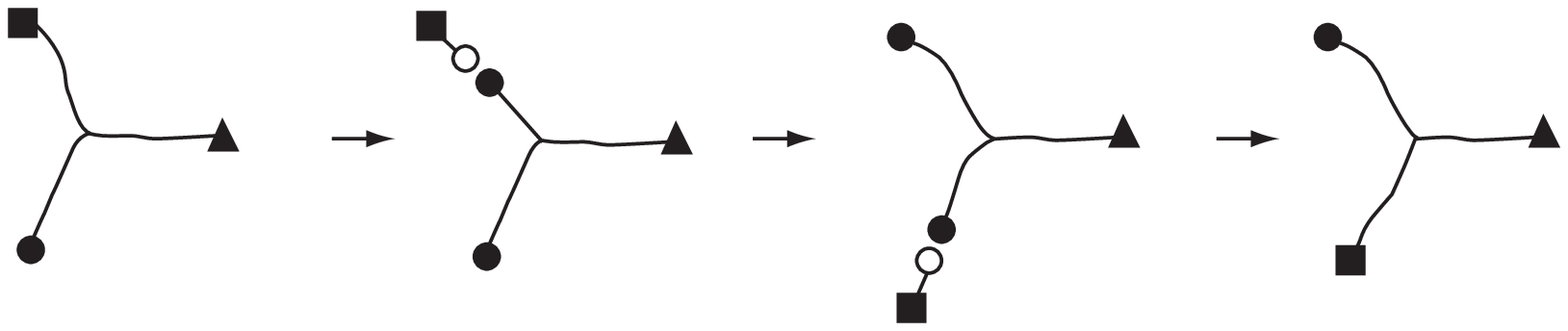}
\vspace*{0.1in}
~
\end{center}

\noindent{Fig. 1(b):  A cartoon of the space-time development of the Z-graph-induced
meson exchange in a baryon in the flux tube model.  For diagrammatic
clarity three different flavors of quarks are shown.  Note that if the
created meson rejoins the flux tube from which it originated, the produced
$q \bar q$ pair can be of any flavor; however, such a process would be a closed $q \bar q$
loop and therefore not part of the quenched approximation.  Also possible,
but not shown, are OZI-violating graphs with the creation or annihilation of
a disconnected $q \bar q$ meson; these are irrelevant to octet meson exchange in
the SU(3) limit and enter in broken SU(3) only through the $\eta-\eta'$
mixing angle.}

\bigskip

	Figure 2 shows what we know about the evolution of quarkonium
spectroscopy as a function of the quark masses.  In heavy quarkonia ($b \bar b$ and
$c \bar c$) we know that hyperfine interactions are generated by one-gluon-exchange
perturbations of wave functions which are solutions of the Coulomb-plus-linear
potential problem.  I find it difficult to look at this diagram and not see
a smooth evolution of the wavefunction (characterized by the slow evolution
of the orbital excitation energy) convoluted with the predicted $1/m_Q^2$
strength of the OGE hyperfine interaction.

\bigskip\bigskip

%
%
\begin{center}
~
\epsfxsize=3.5in  \epsfbox{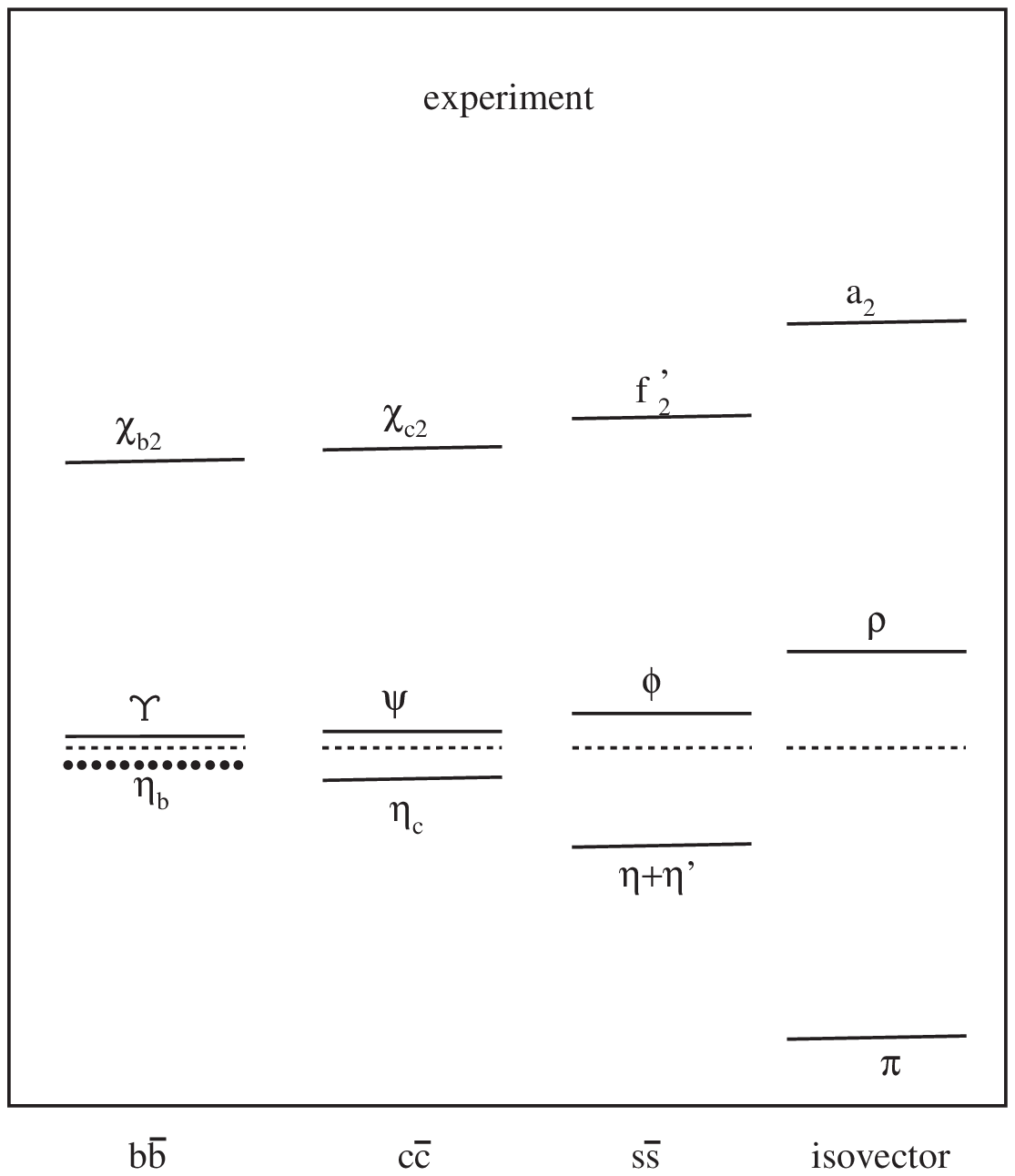}
\vspace*{0.1in}
~
\end{center}

\noindent{Fig. 2: The experimental spectra of $b \bar b$, $c \bar c$, 
$s \bar s$, and isovector light
quarkonia, with the center of gravity of the $S$-wave mesons aligned.  
The $2^{++}$ states have been used to represent the $P$-wave mesons.
The pseudoscalar $s \bar s$ state (``$\eta+\eta'$") has been located by
unmixing a $2 \times 2$ matrix assumed to consist of primordial 
$s \bar s$ and ${1 \over \sqrt 2}(u \bar u+d \bar d)$ 
states.  The $\eta_b$ is not yet discovered, but the theoretical
prediction is shown as a dotted spectral line. The spectra are shown to scale,
which may conveniently be calibrated with the $\chi_{c2}-\psi$ splitting of 459 MeV.}

\bigskip

%
%
\begin{center}
~
\epsfxsize=5.0in  \epsfbox{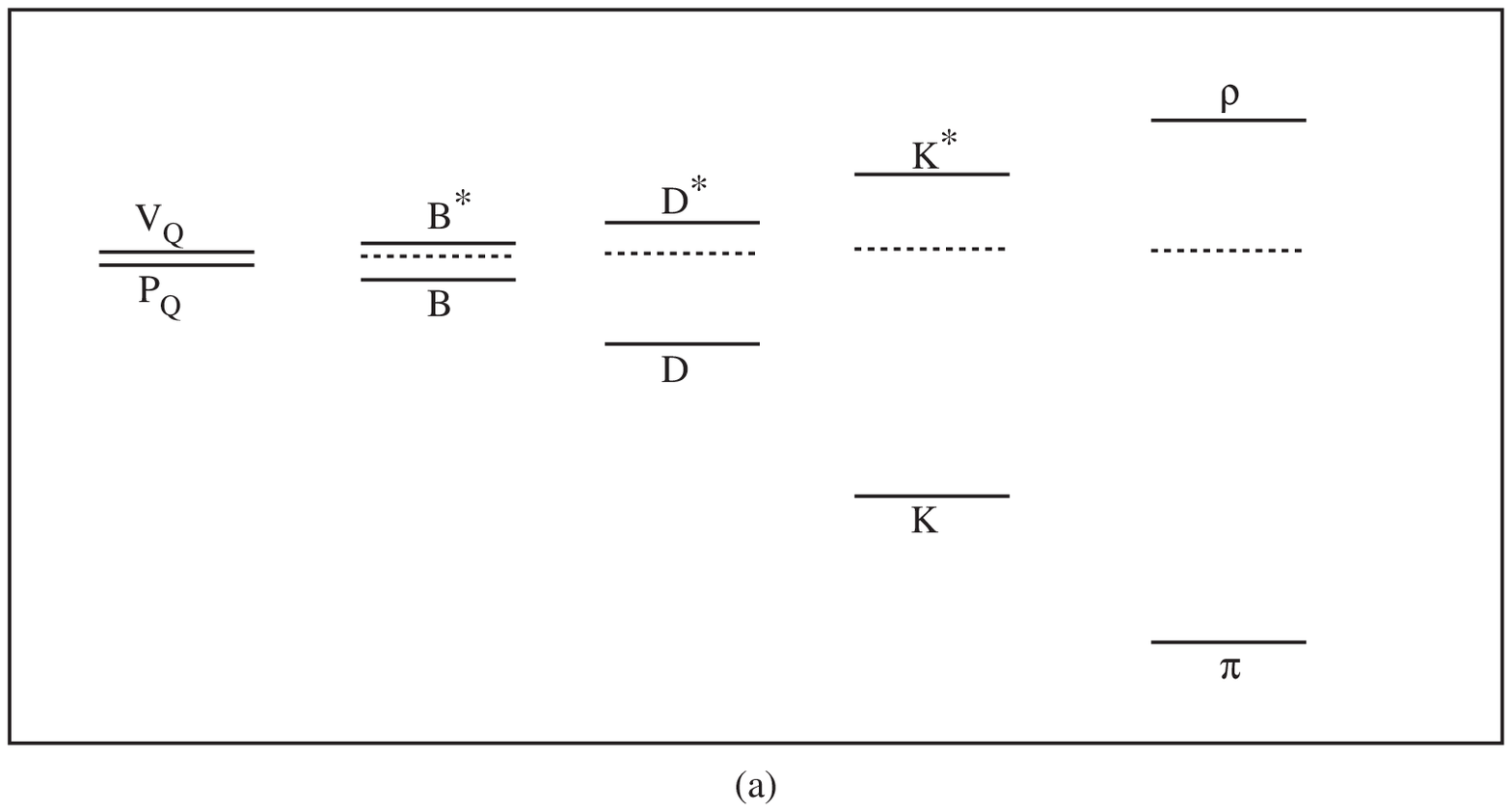}
\vspace*{0.1in}
~
\end{center}

%
%
\begin{center}
~
\epsfxsize=5.0in  \epsfbox{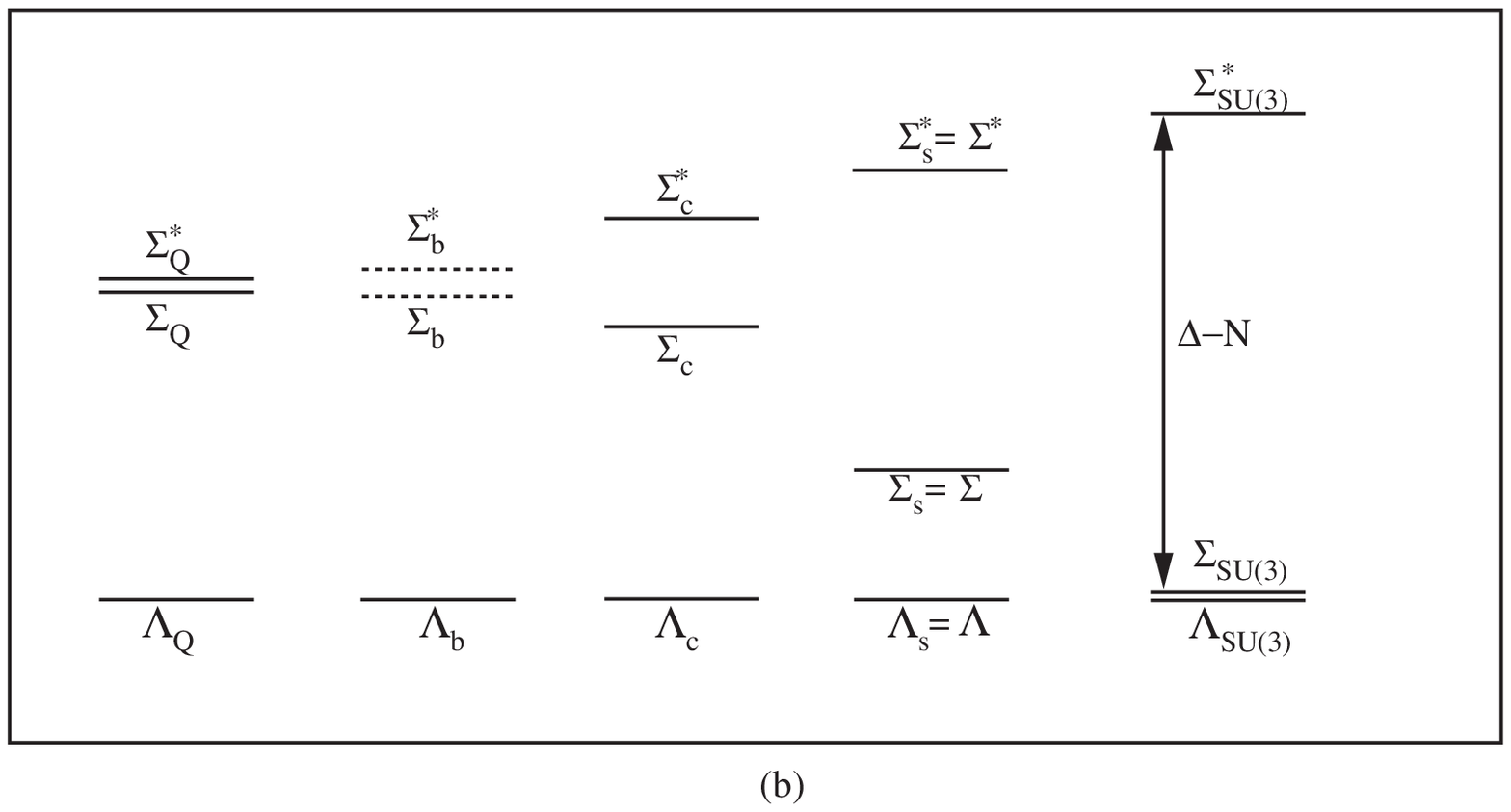}
\vspace*{0.1in}
~
\end{center}

\noindent{Fig. 3: Ground state meson (a) and baryon (b) hyperfine splittings in heavy-light
systems as a function of the mass $m_Q$ of the heavy quark. The spectra on the far left
are the $m_Q \rightarrow \infty$ limits of heavy quark symmetry. The  
$\Sigma^*_Q-\Lambda_Q$ splitting 
and the 
positions of 
$\Sigma^*_b$ and $\Sigma_b$ 
are estimates from the quark model; all other masses are from experiment. 
The spectra are shown to scale; the meson scale
may conveniently be calibrated with the $D^*-D$ splitting of 141 MeV and
the baryon scale with the $\Sigma_c-\Lambda_c$ splitting of 169 MeV.}

\bigskip

	This same conclusion can be reached by approaching the light
quarkonia from another angle.  Figure 3(a) shows the evolution of heavy-light
meson hyperfine interactions from the heavy quark limit to the same
isovector quarkonia.  In this case we know that in the heavy quark limit
\cite{IW} the hyperfine interaction is given by the matrix element of the operator
$\vec \sigma_Q \cdot \vec B/2m_Q$.
In contrast to heavy quarkonium, however, we do not know
that the chromomagnetic field $\vec B$ at the position of $Q$ is being generated by
one gluon exchange from a light valence antiquark.  Nevertheless, by
considering unequal mass heavy quarkonia $Q \bar q$ with $m_q$
beginning at $m_Q$ and decreasing to the light quark mass $m_d$, one finds that
the OGE hyperfine interaction extrapolates very neatly from the end of the
region where it may be rigorously applied ($m_q \simeq 1$ GeV) down to light quark
masses.  The conclusion that heavy-light meson hyperfine interactions are
controlled by OGE is also supported by the striking $1/m_Q$ behaviour of the
ground state splittings in Fig. 3(a)
as $m_Q$ is decreased from $m_b$ to $m_c$ to $m_s$ to $m_d$:
it certainly appears that for all quark masses the quark Q interacts
with $\vec B$ through its chromomagnetic moment $\vec \sigma_Q/2m_Q$, as would be
characteristic of the OGE mechanism \cite{Ds*Ds}.

	     Since the OPE mechanism cannot contribute in mesons, the
OGE mechanism is thus the natural candidate for generating meson hyperfine
interactions.  The objective (as opposed to aesthetic)
problem that arises for the OPE hypothesis
is that it is then nearly
impossible to avoid the conclusion that OGE is also dominant in baryon
hyperfine  interactions:  the OGE $q \bar q$
and $qq$ hyperfine interactions are related by a simple factor of $1/2$, and
given the similarities of meson and baryon structure (for example, their charge
radii, orbital excitation energies, and magnetic moments are all similar),
it is inevitable that the matrix elements of OGE in baryons and mesons are
similar.  Valence quark model calculations support this
qualitative argument, finding that mesons and baryons can be described by a
universal confining potential with one-gluon exchange at short distances
\cite{universal}.

	     There is
another very serious problem with the OPE mechanism which surfaces in
mesons.  I have explained that there are no Z-graph-induced meson exchanges
in mesons.  However,  Fig. 4 shows how the same meson
exchanges which are assumed to exist in baryons 
will drive  mixings in isoscalar channels {\it by annihilation graphs}. More
mechanically, the OPE mechanism posits the existence of vertices by which quarks
couple to pseudoscalar mesons; antiquarks necessarily couple with 
the same strength to the
charge conjugate mesons. By
considering the flavor structure of the allowed vertices, it is easy to show
that the resulting pseudoscalar meson exchange between the quark and
antiquark in a meson must have the character shown in Fig. 4, {\it i.e.}, it
can only operate in isoscalar channels.

\bigskip

%
%
\begin{center}
~
\epsfxsize=1.0in  \epsfbox{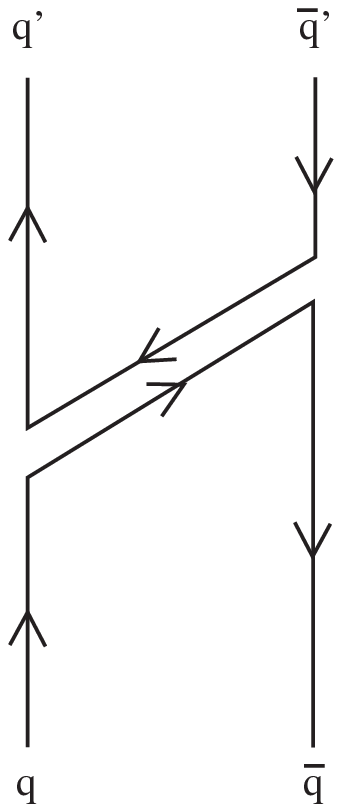}
\vspace*{0.1in}
~
\end{center}

\centerline{Fig. 4:  OZI-violating mixing in isoscalar mesons via the exchange of a $q \bar q'$
meson.}

\bigskip

	I have argued above that the structure of mesons and baryons 
is so similar that it is impossible to avoid
their having similar OGE matrix elements.  The same is true
for OPE matrix elements:  it is impossible to maintain that OPE is
strong enough to produce the $\Delta-N$ splitting in baryons without
predicting a matrix element of comparable strength associated with Fig. 4 in mesons.
Such matrix elements will violate the OZI rule \cite{OZI}. 
Consider the mixing between the pure
$\omega$-like state ${1 \over \sqrt 2}(u \bar u+d \bar d)$ 
and the pure $\phi$-like state $s \bar s$.  This mixing will be
driven by kaon exchange and from the preceding very general arguments we
must expect that the amplitude $A_{OZI}$ for this OZI-violating process will
have a strength of the same order as the 200 MeV  $\Sigma^*-\Sigma$ splitting
(which is also driven purely by kaon exchange). Such an amplitude would be
an order of magnitude larger than that observed:  $A_{OZI}$ 
for the vector mesons is
very tiny - - - of the order of 10 MeV - - - corresponding to the known near purity
of the $\phi$ as an $s \bar s$ state.  The only escape from this disaster is to
argue for some mechanism external to the model which could cancel the large
kaon exchange contribution to $A_{OZI}$.  Given that $A_{OZI}$ is of the order
of 10 MeV in not only the $1^{--}$ mesons, but also in all of the other known
meson nonets (except the pseudoscalars), this escape route seems
implausible.

The mesons thus produce some
disastrous conclusions for the Glozman-Riska model.  The first 
is the very
unaesthetic conclusion that two totally distinct mechanisms are in operation
producing meson and baryon spin-dependent interactions:  OGE in mesons and OPE
in baryons.  The second is the virtual impossibility of having strong OGE
matrix elements in mesons without also producing strong OGE matrix elements
in baryons, in conflict with the basic hypothesis of that model. The third
is that the OPE mechanism produces unacceptably large OZI violation
in meson nonets.

\subsection {The Connection to Heavy Quark Baryons is Lost}

	   As shown in Fig. 3(b), the baryon analog of Fig. 3(a),
experiment provides further strong 
evidence in support of the dominance of OGE and {\it not} OPE in the baryons themselves! 
It is clear from this Figure that in the heavy quark
limit the OPE mechanism is not dominant:  exchange of the heavy
pseudoscalar meson $P_Q$ would produce a hyperfine interaction that scales
with heavy quark mass like $1/m_Q^2$, while for heavy-light baryons the
splittings are behaving like $1/m_Q$ as in the heavy-light mesons.
This is as demanded by heavy quark
theory where these splittings are once again rigorously
controlled by the matrix elements of
$\vec \sigma_Q \cdot \vec B/2m_Q$.

It is
difficult to look at this diagram and not see a smooth evolution of this $1/m_Q$
behaviour from  $m_c$ to $m_s$ to $m_d$, where by SU(3) symmetry
$\Sigma^*_{SU(3)} - \Lambda_{SU(3)}=\Delta - N$, the splitting under
discussion here.  Indeed, using standard constituent quark masses 
($m_d=m_u \equiv m =0.33$ GeV, $m_s=0.55$ GeV, $m_c=1.82$ GeV, $m_b=5.20$ GeV),
the OGE mechanism with its natural $1/m_Q$ behaviour
{\it quantitatively} describes these spectra. In this picture,
$\Sigma^*$, $\Sigma$, and $\Lambda$ are the analogs of
the heavy quark states $\Sigma_Q^*$, $\Sigma_Q$, and $\Lambda_Q$.
I speak of
analogs here because the heavy quark expansion cannot be justified 
for such light values of $m_Q$.
Nevertheless, one expects and observes in both mesons
and baryons the remnants or analogs of heavy quark spectroscopy in light
quark systems.  For example, in Fig. 3(a) the $D^*-D$ heavy quark spin multiplet is
naturally identified with the $K^*-K$ multiplet, {\it i.e.,} the basic degrees of
freedom seen in the spectrum are the same, and from the observed $D^*-D$
splitting and the $1/m_Q$ heavy quark scaling law one expects 
a $K^*-K$ splitting of 460 MeV,
quite close to the actual splitting of 400 MeV.
One might try to escape this conclusion by arguing that
between $m_c$ and $m_s$ the OGE-driven $1/m_Q$ mechanism turns off 
in baryons and the $1/m_Q^2$ OPE
mechanism turns on.  From the baryon spectra alone, one cannot rule out this
baroque possibility.  However, in the heavy-light mesons of Fig. 3(a) there is
no alternative to the OGE mechanism, and if the $Q \bar q$ interaction continues
to grow like $1/m_Q$ as $m_Q$ gets lighter, then (given the
similarity of meson and baryon structure) so must the $Qq$ interaction.  I
see no escape from the conclusion that OGE is  dominant in {\it all} ground
state hyperfine interactions.

The excited
charmed baryon sector has recently provided further strong evidence for
the dominance of the OGE mechanism in baryons. Recall the conclusion
of Section B above that
the $\Lambda (1405){1 \over 2}^-$ and $\Lambda (1520){3 \over 2}^-$ are 
spin-orbit partners.
Heavy quark symmetry \cite{IW} demands that in the heavy-light isospin
zero $\Lambda_Q$ sector, the $\Lambda_Q{1 \over 2}^-$ and $\Lambda_Q {3 \over 2}^-$ be
degenerate as $m_Q \rightarrow \infty$ and that their splitting open up like $1/m_Q$ as $m_Q$
decreases.  The $\Lambda_c(2594){1 \over 2}^-$ and 
$\Lambda_c(2627){3 \over 2}^-$ \cite{PDG}
appear to be such a nearly degenerate pair of states in the charmed baryon
sector.  The center-of-gravity 
${2 \over 3}m_{\Lambda_Q {3 \over 2}^-}  + {1 \over 3}m_{\Lambda_Q{1 \over 2}^-}$
of these two states is 330 MeV above the
$\Lambda_c(2285)$.  This is to be compared with the center-of-gravity of the
$\Lambda (1520)$ and $\Lambda (1405)$ which lies 365 MeV above the $\Lambda(1115)$, 
in accord with the expectation from the quark model that the
orbital excitation energy of the negative parity excitations of $\Lambda_Q$
will be a slowly increasing function of $1/m_Q$.  (The quark model makes a
similar prediction for the P-wave heavy-light mesons which is confirmed by
the data.)  
This alone suggests that the strange quark analogs of the heavy
quark spin multiplet $(\Lambda_Q{3 \over 2}^-, \Lambda_Q{1 \over 2}^-)$ which 
$(\Lambda_c(2627){3 \over 2}^-, \Lambda_c (2594){1 \over 2}^-)$ exemplifies should exist just around
the mass of the $\Lambda (1520){3 \over 2}^-$ and $\Lambda (1405){1 \over 2}^-$.    
Moreover, using the
predicted $1/m_Q$ behaviour of the 
$(\Lambda_Q{3 \over 2}^-, \Lambda_Q{1 \over 2}^-)$ multiplet
splitting would lead to a predicted splitting of 110 MeV in the $\Lambda$
sector compared to the observed splitting of 115 MeV. It is thus very
difficult to avoid identifying 
$(\Lambda (1520){3 \over 2}^-, \Lambda (1405){1 \over 2}^-)$ as
the strange quark analogs of a heavy quark spin multiplet, and
to avoid concluding that the $1/m_Q$ evolution of the OGE mechanism is responsible
for its splitting.

	    Heavy quark symmetry thus poses another serious problem for the Glozman-Riska model.
The  OPE mechanism not only fails to explain meson
hyperfine interactions, but it also cannot even explain the spectra of all baryons:
it violates the requirement that splittings in
heavy-light baryons open up like $1/m_Q$.
Heavy-light mesons show that the $1/m_Q$ behaviour of the OGE mechanism
persists all the way down to light quark masses, and the observed behaviour
of baryons is completely consistent with the same extension of the heavy quark
symmetry scaling laws. This behaviour leaves very little room for the OPE
mechanism, and certainly makes it implausible that it is dominant.

\section {Conclusions}

	I have focused in this paper on the predictions of the
Glozman-Riska OPE model \cite{GR}.  I believe the
catalogue of problems I have described are
sufficient to 
rule the model out.  

    Though in its extreme version the model is unsustainable, 
some elements of the physics of the Glozman-Riska model surely play a role
in baryons. While the process depicted in Fig. 1(a) may be taken into account
in a ``dual approximation" by relativistic valence quark propagators,
quark-antiquark correlations in the intermediate state will
produce departures from this approximation. We may expect that
such departures will be most pronounced in situations where the meson spectrum
most strongly breaks the closure limit required for duality \cite{NIadiabatic,GIonV,GIonOZI},
and the pion certainly has the potential to do this. Of course, given that
these exchanges occur at short distances and that the structure of the 
Goldstone bosons is very similar to that of other mesons \cite{GBordinary},
there is no obvious rationale for truncating the tower of meson exchanges associated
with Fig. 1 with these states alone. (I hasten to add that it is {\it not}
the distance to threshold which directly determines the importance of a given meson:
its dominant contribution comes from the peak of its spectral function, and 
in realistic models this
feature is controlled more by the internal structure of a 
meson than by its mass \cite{NIadiabatic,GIonV,GIonOZI}.)

    The $1/N_c$ expansion offers additional insights into this issue
and into the structure of the arguments presented in this paper. Since
OGE and OPE are of the same order in $1/N_c$ in baryons, there is unlikely to be any
general principle which could be used to decide which is dominant. This situation can
be contrasted with the case of heavy quarks for which the Z-graphs of Fig. 1(a) are
suppressed like $\Lambda_{QCD}/m_Q$ relative to gluon exchange graphs. The dominance
of OGE over OPE in baryons must therefore have a dynamical origin. In contrast to
baryons, OPE-like effects in mesons are suppressed by a power of
$1/N_c$ relative to gluon exchange. One can thus look to mesons for a 
relatively unobscured picture of
the strength and character of OGE effects, as I did  in many of the
arguments of this paper.

    It remains to speculate on why $q \bar q$ pair creation effects are not more
important in baryons. The OPE (or more generally meson-exchange) potentials between
quarks would only be the simplest manifestation of such effects. More generally, meson
emission and reabsorption by the baryons would lead to a complex interaction of the
discrete baryon spectrum with each of the baryon-meson continua. Studies of
``unquenching the quark model" have provided a plausible explanation for why such
effects do not demolish the spectroscopy of the quark potential models
\cite{NIadiabatic,GIonV} and the success of the OZI rule \cite{GIonOZI}. These studies
indicate that the resiliency of valence quark model spectroscopy to $q \bar q$ pair
creation occurs {\it not} because such processes are intrinsically weak, but because
most of their effects can be absorbed into renormalized valence quark model parameters.
The prime example of this absorption of $q \bar q$ effects is the string tension.
Though the string tension is a strong function of the number of light flavors,
when it is carefully renormalized to its observed value
it produces the observed spectrum. A particularly treacherous aspect of
this situation is that if one examines the effect of any particular continuum channel
on the spectrum, it may appear to be very large and to have a complex dependence on
the state of the quarks; only the sum over all channels leads in first
approximation to a simple renormalization of the string tension. After this
summation, only small residual effects associated with nearby thresholds remain 
\cite{NIadiabatic}.  

   The renormalization of the string tension is associated
with $q \bar q$ pair
creation in which the pair forms a closed loop. However, a created $q \bar q$ pair of
the appropriate flavor could also join with a valence quark or antiquark to make a
Z-graph. As already mentioned, the sum over all such processes at the hadronic level is
apparently dual to the leading Z-graph component of the relativistic valence quark
propagator. In baryons, meson exchange between quarks is contained in the Z-graph
process of Fig. 1(a), and therefore the bulk of the effect of such hadronic
processes should be absorbed into renormalized valence quark model parameters,
in this case those describing the relativistic valence quark propagator.   
I speculate that duality is once again sufficiently
accurate that only a small residue of this second type of $q \bar q$ effect  
remains after summing over
channels, suppressing meson exchange
relative to its naive strength in a $1/N_c$ expansion.

   It would be interesting to apply the methods of Refs. \cite{NIadiabatic,GIonV,GIonOZI}
to the Z-graphs to see if the effects of duality violation are indeed small.
However, whatever the reason, the empirically-based arguments of this
paper make it clear that gluon exchange and not meson exchange
provides the dominant residual forces between constituent quarks in both mesons and baryons.

\vfill\eject

{\centerline {\bf ACKNOWLEDGEMENTS}}

\medskip

    This work was supported by DOE contract DE-AC05-84ER40150 under which the
Southeastern Universities Research Association (SURA) operates
the Thomas Jefferson National Accelerator Facility.

\bigskip\bigskip

{

\end{document}
\begin{references}
\bibitem{Glozmanreply} A quite different opinion is expressed
in L. Ya. Glozman, nucl-th/9909021, which is a reply to the first
version of this paper. This version has been modified to further reduce
the discussion of practical details and theoretical 
foundations, to add a number of references
that have been brought to my attention, and to clarify a few
arguments; the  catalogue of criticisms remains unchanged.
\bibitem{3/2photo} This example was also selected because it is the
most prominent case out of several
hundred where
the  decay amplitude analyses have made trouble for the Isgur-Karl model.
While as shown in Ref. \cite{KoniukI}
the internal structures of the five $N^*{3 \over 2}^+$ states are consistent
with only the
$N^*{3 \over 2}^+(1720)$ having been observed, its photocouplings cast
doubt on the association
of this state with the predicted state of the Isgur-Karl model.
\bibitem{Ds*Ds} I omit the $D_s$ and $\bar B_s$ systems from this diagram so that one
might view the effect of changing $m_Q$ while holding the light quark mass fixed at
$m_d$. Under these conditions, it is predicted that the mean radius of the wave
functions varies across the whole diagram by only $25 \%$ so that the $1/m_Q$ effect is
dominant. In going from the $D$ to the $D_s$ system or the $\bar B$ to $\bar B_s$
system, the wave function is in contrast predicted to change so strongly as to
completely compensate for the reduced strange quark hyperfine interaction, so that
the effect we are trying to visualize is obscured.
\end{references}
